# Symplectic integration approach for metastable systems

E. Klotins

*Institute of Solid State Physics, Riga, Latvia*

Nonadiabatic behavior of metastable systems modeled by anharmonic Hamiltonians is reproduced by the Fokker-Planck and imaginary time Schrödinger equation scheme with subsequent symplectic integration. Example solutions capture ergodicity breaking, reassure the H-theorem of global stability [M. Shiino, Phys. Rev. A, Vol 36, pp. 2393-2411 (1987)], and reproduce spatially extended response under alternate source fields.

## I. INTRODUCTION

Metastable systems is an active field of study steamed by developments in condensed state physics with model Hamiltonians and relevant Fokker-Planck equations as natural framework of the theory [1]. However, its capacity to reproduce critical phenomena crucially depends on the mathematical technique and in this context the Wentzel-Kramer-Brilluin (WKB) analysis based on mapping between Fokker-Planck and imaginary time Schrödinger equation has received a renewed attention [2] and application to linear problems [3,4]. The subject of this work is nonlinear Fokker-Planck – imaginary time Schrödinger equation scheme emerging for metastable systems exhibiting critical behavior. Special attention is paid to polarization response in ferroelectrics modeled by Ginzburg – Landau type model Hamiltonians. Unlike the customary direct integration of kinetic equations [5,6] the aforementioned scheme yields a systematic solution of the Dirichlet problem. The paper is organized as follows. In Sec.II we give insight in the Fokker-Planck – imaginary time Schrödinger equation scheme and its application to dynamic hysteresis [7,8]. In Sec.III this scheme is extended for the model of globally coupled anharmonic oscillators reproducing polarization response under alternate source field. In Sect.IV the spatial extension and finite size effects are reproduced within the model of locally coupled anharmonic oscillators. The physical background as well as key problems for further developments is analyzed in Sect.V.

## II. FOKKER-PLANCK AND IMAGINARY TIME SCHRÖDINGER EQUATION SCHEME

The starting point for further generalization is the Fokker-Planck equation for probability density of the order parameter (polarization) $\rho(P,t)$

$$\frac{\partial \rho(P,t)}{\partial t} = \frac{\partial}{\partial P}\left(\frac{\delta U}{\delta P}\rho(P,t)\right) + \Theta\frac{\partial^2 \rho(P,t)}{\partial P^2} \qquad (1)$$

The (dimensionless) energy functional

$$U = -P^2/2 + P^4/4 + (\nabla P)^2/2 - \lambda(t) P, \qquad (2)$$

accentuated in this section emerges from arbitrary model Hamiltonian $H = \int U(\mathbf{P})dV$ and the integration over volume is ignored because it is not essential for further approach. The diffusion coefficient $\Theta$ comes out from the microscopic degrees of freedom as well as from the prefactor $\gamma$ of corresponding Langevin equation (in physical units) $\frac{\partial P(\mathbf{x},t)}{\partial t} = -\gamma\frac{\delta U[P(\mathbf{x},t)]}{\delta P(\mathbf{x},t)} + \eta(t)$. Customary the dependence of the diffusion coefficient $\Theta$ on other parameters of theory including temperature is assumed as noncritical. The factor $\lambda(t)$ denotes alternate driving (source) field, and the gradient term $(\nabla P)^2$ (omitted in this section for mathematical convenience) specify first neighbor interaction. The concept is to transform Eqs.(1,2) in imaginary time Schrödinger equation for the auxiliary function $G(P,t)$ introduced by the standard WKB ansatz [2] and unfolding expectation value of polarization through the first moment of probability density $\rho(P,t)$

$$\rho(P,t) = \exp[F(P)]G(P,t) \qquad (3)$$

The imaginary time Schrödinger equation reads as

$$\frac{\partial G(P,t)}{\partial t} = \left[\Theta\frac{\partial^2}{\partial P^2} + V(P)\right]G(P,t) \qquad (4)$$

and the potential operator $V(P)$ is given by

$$V(P) = \left[-\frac{1}{4\Theta}[U'(P)]^2 + \frac{1}{2}U''(P)\right] \qquad (5)$$

The survey includes analytical solution of an ordinary differential equation for $F(P)$ in Eq.(4) canceling the first derivative of auxiliary function in Eq.(4) and simultaneously determining the WKB ansatz as

$$\rho(P,t) = \exp[-U(P)/2\Theta]G(P,t) \qquad (6)$$

The mapping between Eqs.(1,4) is quite general and applicable for arbitrary energy functionals. The analytical and quite exact part of computations is completed by recurrence relation for the auxiliary function valid for a small time slice $\Delta t$

$$G(P, t + \Delta t) = \exp\left[\Delta t \left(\Theta \frac{\partial^2}{\partial P^2} + V(P)\right)\right] G(P, t) \qquad (7)$$

The symplectic integrator for Eq.(7) reads as

$$\left(1 - \frac{\Theta \Delta t}{2} \frac{\partial^2}{\partial P^2}\right) G(P, t + \Delta t)$$
$$= \left\{\exp\left[\frac{\Delta t}{2} V + \frac{\Delta t^3}{48} (\nabla V)^2\right]\left(1 + \frac{\Theta \Delta t}{2} \frac{\partial^2}{\partial P^2}\right)\exp\left[\frac{\Delta t}{2} V + \frac{\Delta t^3}{48} (\nabla V)^2\right]\right\} G(P, t) \qquad (8)$$

and the potential operator $V$ is given with time argument $t := t + \Delta t / 2$ [9,10] so accounting for time dependence of the model Hamiltonian. Symplectic integration Eq.(8) is motivated by its norm conservation and long term stability and, as showed hereafter, applicability to ergodicity breaking and bifurcation of relaxation time as a natural extension of Eqs.(1-8).

A simplest test solution for the symplectic integration concern dynamic hysteresis - the combined effect of periodic source field and additive noise - exhibiting, in the absence of external source, a unique ground state. Dynamic hysteresis plots [3] based on quartic energy functional are in agreement with analytical results [7]. Application of the symplectic integration technique to $(\Psi^2)^3$ type energy functional $U(P_1) = \alpha_1 P_1^2 + \alpha_{11} P_1^4 + \alpha_{111} P_1^6$ within the range of metastability is demonstrated in Fig.1. The parameters for $PbTiO_3$ [11] are as follows: Curie-Weiss constant $1.5\,10^5\ ^0C$, transition temperature $T_C = 492.2$ ($^0C$), $\alpha_1 = 61 \cdot 10^5$ (m/F) at $T_C$, $\alpha_{11} = -9.235 \cdot 10^7$ (m$^5$/(C$^2$F), $\alpha_{111} = 3.469 \cdot 10^8$ (m$^9$/(C$^4$F).

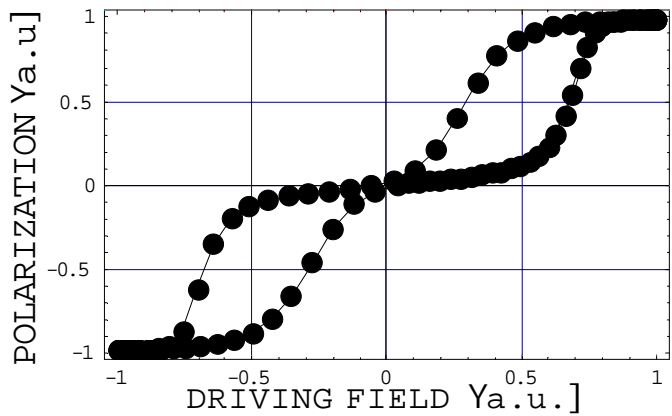

Fig.1 Dynamic hysteresis plot for $PbTiO_3$ at temperature $T = 1.005\,T_C$. The source field is given in units of a $0.15$ maximum value at which the energy landscape transforms in a single wall one. The rest of parameters are as follows: diffusion constant $1/100$, frequency $10^{-3}$. The polarization is normalized by the spontaneous one.

Another test solution demonstrates matching between the phenomenological and first principles calculations [8] in which the nonlinear dielectric and piezoelectric response of tetragonal $PbTiO_3$ is modeled by energy functional $F(\eta_1, \eta_3, P_z)$ (in denominations of [8])

$$F = \frac{1}{2}c_{11}(2\eta_1^2 + \eta_3^2) + c_{12}(2\eta_1\eta_3 + \eta_1^2) + A_2 P_z^2 + A_4 P_z^4 + A_6 P_z^6 + 2B_{1yy}\eta_1 P_z^2 + B_{1zz}\eta_3 P_z^2 \quad (9)$$

Here $\eta_1 = \eta_{xx} = \eta_{yy}$ and $\eta_3 = \eta_{zz}$ are components of strain tensor, and the expansion is truncated to the first order in elastic and polarization-strain coupling. Clamped-strain response obtained by zeroing the variations of Eq.(9) with respect to the components of strain tensor yields renormalization of the expansion coefficient at $P^4$ in Eq.(9) as follows[8]

$$F(P_z) = A_2 P_z^2 + \left(A_4 + \frac{2c_{12}B_{1zz}B_{1yy} - c_{11}B_{1yy}^2 - (c_{11} + c_{12})B_{1zz}^2/2}{(c_{11} + 2c_{12})(c_{11} - c_{12})}\right) P_z^4 + A_6 P_z^6 \quad (10)$$

Since the first expansion coefficient $A_2 < 0$ the energy landscape exhibit two local minima resulting in static hysteresis loop as shown by line in Fig.2. Numerical values of the parameters for $PbTiO_3$ [8] are as follows: $A_2 = -0.003$, $A_4 = 0.005$, $A_6 = 0.004$ (in the units of $Ha$ and $C/m^2$) and the components of elastic tensor and the coefficients of polarization-strain coupling are $c_{11} = 4.374$, $c_{12} = 1.326$, $B_{1zz} = -1.99$ and $B_{1yy} = -0.049$, correspondingly. Modeling of dynamic hysteresis within the scheme Eqs.(1-8) starts with variation of Eq.(10) and yields Langevin equation (in physical units)

$$\frac{\partial P_z}{\partial t} = -\gamma \frac{\delta[F(P_z) - P_z\lambda(t)]}{\delta P_z} + \eta(t) \quad (11)$$

Here the parameters are the prefactor (kinetic coefficient) $\gamma$, period of source field, and the noise strength $\eta(t)$. At appropriate fitting (diffusion coefficient $1/1000$, kinetic coefficient $\gamma = 10$) the dynamic hysteresis plot (dots) match fairly well with the static one.

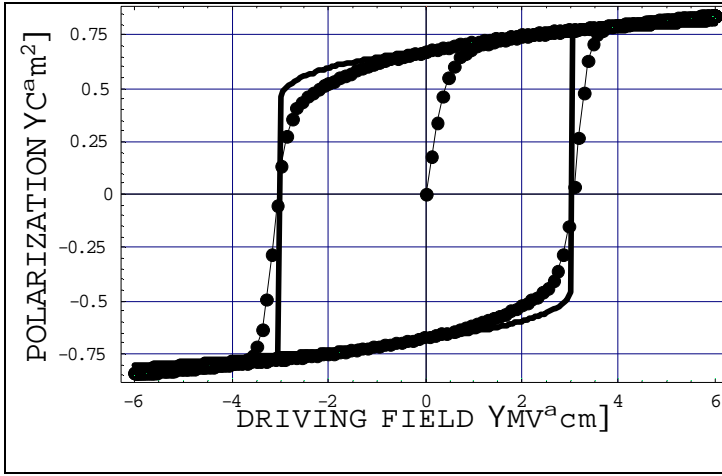

Fig.2 Comparision of first principle clamped-strain static hysteresis plot (bold line) [8] and the polarization response on periodic source (dots). At high frequency source field ($\Omega = 10^{-2}$) the dynamic hysteresis plot approaches to the first principle one and the coercive field resembles the static one, as expected.

### III. MODEL OF COUPLED ANHARMONIC OSCILLATORS

Customary Ginzburg – Landau type models reproduces thermodynamic equilibrium properties by solving a variational problem in infinite volume limit. Depending on the temperature the system remains ergodic or experiences ergodicity breaking and bifurcation of the spontaneous polarization. However, for statistical approach formally given in terms of Langevin equations and accounting for noise source (bath) [12,13] these features are lost unless a more complex model is introduced. In this context the model of coupled overdamped anharmonic oscillators [14] receives attention. This description requires that it is possible to divide the spatial domain in a large number of cells in such a way that each cell is small enough for all the oscillators in the cell can be assumed to possess the same characteristics of the cell. A set of Langevin equations for the polarization $P(t)$ reads as

$$\frac{\partial P_i}{\partial t} = -\frac{\partial F(P_i)}{\partial P_i} + \sum_{k=1}^{N} \frac{\varepsilon}{N}(P_k - P_i) + \eta_i(t) \qquad (12)$$

here the stochastic terms $\eta_i(t)$ determines the simplest bath specified by statistically independent white noise with correlation function $\langle \eta_i(t)\eta_j(t') \rangle = \delta_{ij}\delta(t-t')$ and the factor $\varepsilon > 0$ denotes the strength of attractive mean-field type coupling. At the thermodynamic $N \to \infty$ limit the averages of $P_k$ in Eq. (12) can be assumed to behave in a deterministic way, namely, $\lim_{N \to \infty}\left(\frac{1}{N}\sum_{k=1}^{N} P_k(t)\right) = \overline{P}(t)$ and the corresponding Fokker-Planck equation concern probability density for each $P_i$ which originates from various realizations of white noise

$$\frac{\partial \rho}{\partial t} = \sum_{i=1}^{N}\left[-\frac{\partial}{\partial P_i}\left[-\frac{\partial F}{\partial P_i} + \frac{\varepsilon}{N}\sum_{k=1}^{N} P_k - \frac{\varepsilon}{N}\sum_{k=1}^{N} P_i\right]\rho + \Theta\frac{\partial^2 \rho}{\partial P_i^2}\right] \qquad (13)$$

Recognizing that $\frac{\varepsilon}{N}\sum_{k=1}^{N} P_k(t) = \varepsilon\overline{P}(t)$ and each $i$-th entity concerns a coarse-grained block described by equal kinetics Eq.(12) the Eq.(13) reduces to the model of globally coupled anharmonic oscillators

$$\dot{\rho}(P,t) = \frac{\partial}{\partial P}\left[U'(P,t) + \Theta\frac{\partial}{\partial P}\right]\rho(P,t) \tag{14}$$

which generates stationary solutions of Eq.(14) exhibiting both bifurcation of the ground state (spontaneous polarization) and divergence of the relaxation time for a rich scale of model Hamiltonians. Stationary solution for simplest prototype energy functional

$$U(P,t) = -\frac{P^2}{2} + \frac{P^4}{4} - P\lambda(t) + \frac{\varepsilon}{2}\left[P - \overline{P}(t)\right]^2 \tag{15}$$

is given in [3]. This stationary state is in agreement with the Boltzmann's H-theorem for global stability (ensuring the existence of a uniquely determined long-time probability distribution $\rho_\infty(P,t)$) and, in case of nonlinearity [14], stating that (at overcritical interaction constant) the system always reaches global stability in the sense that any time dependent solution of Eq.(14) lying far from equilibrium must be attracted by either one of those stationary solutions without any possibility of runaway behavior or limit cycle type oscillations. Temporal response going beyond this theorem is reproduced in the course of symplectic integration and starts with the ansatz Eq.(3) that yields relation for the auxiliary function Eq.(4) $V(P,t) = V_1(P,t) + V_2(\overline{P}(t),P,t)$ made up of both the linear $V_1(P,t)$ and the nonlinear $V_2(\overline{P}(t),P,t)$ terms in the potential operator and explicitly given in [12]. Example solution for a system with initially positive remnant polarization affected by a negative source is demonstrated in Fig.5. The source is modeled by sow tooth shaped variable length pulse. What is anticipated at $t \to \infty$ limit is approaching the expectation value $\overline{P}$ to a remnant polarization, either $P_r$ or $-P_r$. The sign of the expectation value $\overline{P}$ at the time instant $T$ at which the source turns to zero is crucial, namely, at $\overline{P}(t=T) > 0$ the remnant polarization approaches to $\overline{P}(t=\infty) \to P_r$, and $\overline{P}(t=\infty) \to -P_r$ otherwise as it follows from the H-theorem of global stability [14]. This behavior is confirmed in Fig.5 with $\overline{P}(T) = 0$ as the point splitting the $\overline{P}$ - space in two domains of attraction for $P_r$ and $-P_r$. However, time propagation of the system at $0 < t < T$ within which the source field is nonzero goes beyond the H-theorem of global stability [14] and is revealed in the course of symplectic integration.

In the example solution the source field is specified by $-0.027$ (a.u.) amplitude (corresponding $\sim 0.07$ of the thermodynamic coercive field in physical units) and the length varying between $320 \leq T \leq 1580$ a.u. As shown in Fig.5 the $710$ and $1580$ (a.u.) source pulses are obviously overcritical and belong to the $-P_r$ domain of attraction. Otherwise, the $320$ and $640$ (a.u.) source pulses are undercritical and belong to the $P_r$ domain of attraction.

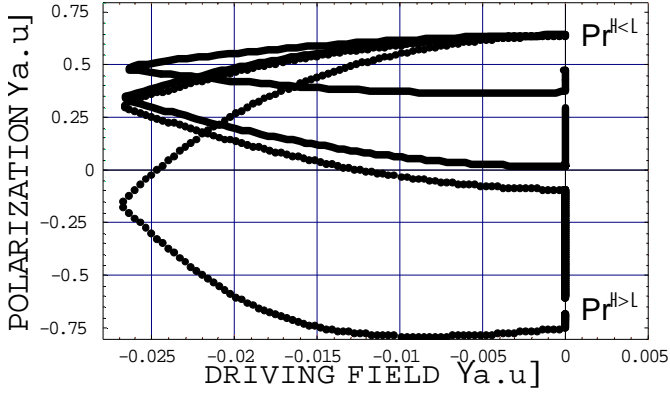

Fig. 3 Temporal response in case of global coupling. Lines illustrate the effect of source pulses being to short (undercritical) for polarization switching and the system remains in the $P_r^{(+)}$ domain of attraction. Otherwise, under pulses with overcritical length (dots) the system enters in the $P_r^{(-)}$ domain of attraction.

Nonlinearity of this problem is managed by guest function method [12] that has essential consequences for spatially extended problem emerged by locally coupled model in which each anharmonic oscillator is coupled which its first neighbors.

## IV. POLARIZATION RESPONSE IN THE MODEL OF LOCALLY COUPLED ANHARMONIC OSCILLATORS

Spatial dependence of polarization field, disappearing in the model of globally coupled anharmonic oscillators Eqs.(12), is restored in the case of first neighbors coupling. This approach assumes that (i) the system consist of finite number of (microscopically large) blocks modeled by Ginzburg-Landay energy functional $\Phi_i = -\frac{1}{2}P_i^2 + \frac{1}{4}P_i^4 - \lambda(t)P_i$ (here we consider the $T < T_C$ case and the sixth order term is irrelevant) and (ii) the first neighbor interaction between (macroscopically small) blocks holds so addressing the problem to ensemble of interacting blocks. Going around the microscopic interpretation of the strength of interaction and the correlation length, the problem is formulated by the model Hamiltonian

$$H \equiv \sum_i^N \left\{ \Phi_i + \frac{\varepsilon}{2}\left( \left(\overline{P}_{i+1}(t) - P_i\right)^2 + \left(\overline{P}_{i-1}(t) - P_i\right)^2 \right) \right\} \quad (16)$$

Here the expectation values $\overline{P}_k$ are unknown quantities and are evaluated selfconsistently afterward. Kinetic equations derived from Eq.(16)

$$\frac{\partial P_i}{\partial t} = -\frac{\partial \Phi_i}{\partial P_i} + \varepsilon\left(\overline{P}_{i+1}(t) - 2P_i + \overline{P}_{i-1}(t)\right) \quad (17)$$

readdress the problem to a set of Fokker-Planck equations for probability distribution

$$\dot{\rho}(P_i,t) = -\frac{\partial}{\partial P_i}\left[ -\frac{\partial \Phi_i}{\partial P_i}\rho(P_i,t) + \varepsilon\left(\overline{P}_{i+1}(t) - 2P_i + \overline{P}_{i-1}(t)\right)\rho(P_i,t) \right] + \Theta_i \frac{\partial^2}{\partial P_i^2}\rho(P_i,t) \quad (18)$$

For illustration purpose we consider 1-d Dirichlet problem for two $180^0$ domains. Initial conditions for probability density of polarization are given by stationary solution of Eq.(18)

$$\rho(P_i) = C \exp\left[\frac{-\Phi(P_i) + \varepsilon P_i(\overline{P}_{i-1} - P_i + \overline{P}_{i+1})}{\Theta_i}\right] \tag{19}$$

Here $C$ is normalization constant, and $\overline{P}_0 = 0$, $\overline{P}_{i_{max}+1} = 0$ are zero boundary conditions. Implementing normalization of the probability distribution as well as the first moment $\overline{P} = \int_{-\infty}^{\infty} P\rho dP$ the selfconsistency condition for $\overline{P}_i$ is given by

$$\frac{1}{C}\left(\int_{P_{min}}^{P_{max}} P_i \exp\left[\frac{-\Phi(P_i,0) + \varepsilon P_i(\overline{P}_{i-1} - P_i + \overline{P}_{i+1})}{\Theta_i}\right]dP_i\right) - \overline{P}_i = 0 \tag{20}$$

Here $C$ is normalization constant [12]. Initial state of Cauchy problem is given by stationary solution of Eq.(18) for starting values $\overline{P}_i^{(s)} = 1$ for $i \in [1, i_{max}/2)$ and $\overline{P}_i^{(s)} = -1$ for $i \in (i_{max}/2, i_{max}]$ as demonstrated in Fig. 4.

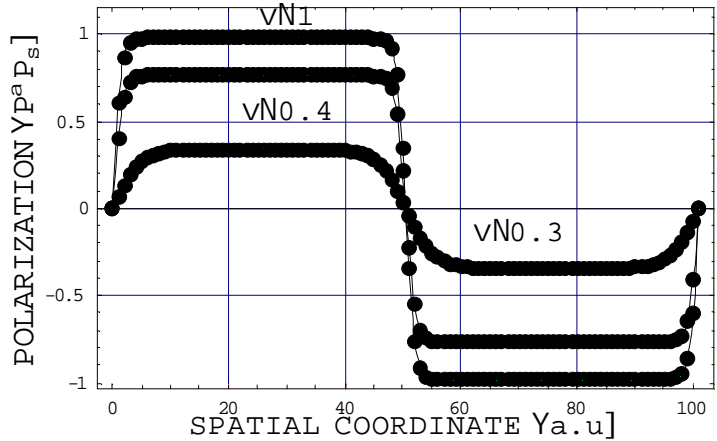

Fig. 4 Stationary solution for $180^0$ domains in a 1-d region with zero boundary conditions and various coupling constants.

Nonstationary solution of Eq.(18) starts with Eq.3 and Fokker-Planck equations Eqs. (22) and yields imaginary time Schrödinger equation for the auxiliary function defined over spatial mesh $i \in [1, i_{max}]$

$$\dot{G}(P_i, t) = [T[i] + V_1[i] + V_2[i] + K[i]]G(P_i, t) \tag{21}$$

Here the kinetic operator $T[i]$ is given by $T[i] = \Theta \frac{\partial^2}{\partial P_i^2}$, the linear part $V_1[i]$ and the nonlinear part $V_2[i]$ of potential operators in Eq.(28) are given by relations $V_1[i] = -\frac{1}{4\Theta_i}\left(\frac{\partial \Phi_i}{\partial P_i}\right)^2 + \frac{1}{2}\left(\frac{\partial^2 \Phi_i}{\partial P_i^2}\right)$ and

$$V_2[i] = \frac{\varepsilon}{4\Theta_i}\left(4\Theta_i - (2P_i - \overline{P}_{i-1}(t) - \overline{P}_{i+1}(t))\left(2P_i - \varepsilon(\overline{P}_{i-1}(t) + \overline{P}_{i+1}(t)) + 2\frac{\partial \Phi}{\partial P_i}\right)\right)$$ ,correspondingly.

Finally, correction to the potential operators generated by explicit time dependence of the energy functional yields as $K[i] = \left[-\varepsilon P_i \vec{P}_{i-1}(t) - \varepsilon P_i \vec{P}_{i+1}(t) + \vec{\Phi}_i\right](2\Theta)^{-1}$. Subsequent numerical calculations include the solution of Eq.(28 21) and evaluation of the merit $M(Q) = \int P\rho(P,Q,t)dP - (\overline{P}(0) + Q\Delta t)$ (transformed in analytical function of $Q_i$ by quadratic interpolation). This trick generates a set of coupled algebraic equations $M(Q_i) = 0$ for expansion coefficients $Q_i$ so returning the density distributions by Eq.(26 3) over spatial mesh in every time slice. Preliminary results of the 1-d domain switching are shown in Fig.5 for a couple of $180^0$ domains Fig.4 assigned as $\overline{P}(0)$. The switching is initiated by a negative source field $\lambda(t) \sim \left[\exp[-(t/150)^2] - 1\right]$ that deviate the value of polarization in both domains. It must be emphasized that the deviation $\overline{P}(t) - \overline{P}(0)$ being negative for any spatial coordinate prevails at the boundaries and at the domain wall in accord with recent estimates [16].

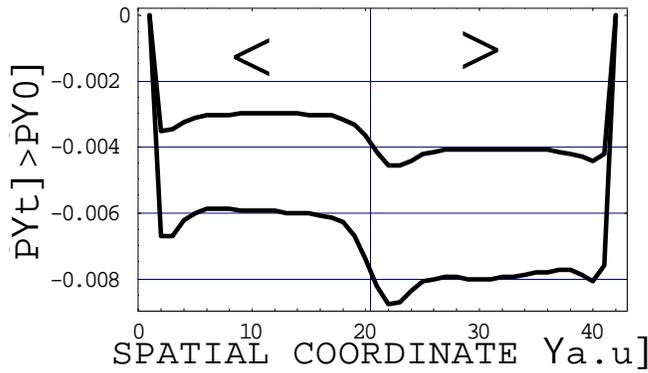

Fig.5 Effect of negative source field at $t = 26$ and $t = 33$ time instants corresponding to $0.007$ and $0.012$ of the thermodynamic coercive field. Representative parameters of the problem: first neighbor coupling constant $\varepsilon = 0.035$, diffusion constant $\Theta = 0.05$.

In 3-d case the quantity of interest is 3-d $G(\mathbf{P}, \mathbf{x}, t)$ − function which inherits the impact of all components of the polarization vector. However, the general structure of symplectic integration scheme Eqs.(1-8) remains unchanged and its spatial extension is straightforward.

## V. DISCUSSSION AND CONCLUSIONS

Symplectic integration of nonlinear Fokker-Planck equations capturing Landau-type critical dynamics motivates the analysis of where do the (over)simplified model Hamiltonians constituted from a set of anharmonic oscillators (blocks) stand. Firstly, for globally coupled systems introduced for mathematical convenience, typically each lattice site is connected to all others with the same coupling strength. Physically meaning are microscopically large and macroscopically small objects within which the order parameter is uniform and obey Landau relations. Mutual effect of coupling and noise reproduces both the ordered phase and the ergodicity breaking. Another level of coarse graining emerges as represented by diffusively coupled blocks located at the sites of a lattice with just nearest neighbor coupling. Formally it yields a generalization of usual thermodynamics for spatially inhomogeneous situations, where the order parameter become a coordinate dependent field $\mathbf{P}(\mathbf{x})$ if smoothed over blocks whose center point lies at $\mathbf{x}$ [1]. An assumption hidden in aforementioned Langevin - Fokker-Planck scheme is that coupling of the prefactor $\gamma$ with the order parameter and other quantities of the theory is not critical.

Nevertheless, it is shown how the Langevin, Fokker-Planck and imaginary time Schrödinger equation techniques can be derived elegantly in terms of symplectic integration even for nonlocal and hardly nonlinear problems and can be used in calculations of response properties of spatially inhomogeneous metastable systems.

**Acknowledgements**

This work has been partially supported by the Contract No. ICA1-CT-2000-70007 of European Excellence Center of Advanced Material Research and Latvian Science Project Nr.01.0805.1.1.